\documentclass{article}
\usepackage[dvips]{graphicx}
\usepackage{latexsym}
\usepackage{amssymb}
\newcommand{\tr}{{\rm tr}}
\newcommand{\R}{{\bf R}}
\newcommand{\next}{\hfil\break\noindent}
\newcommand\f[2]{\textstyle\frac{#1}{#2}}
\renewcommand\d{\partial}

\begin{document}
\date{}

\title {Dynamics of spatially homogeneous solutions of the Einstein-Vlasov 
equations which are locally rotationally symmetric}

\author{A. D. Rendall\footnote{
Max-Planck-Institut f\"ur Gravitationsphysik,
Schlaatzweg 1,
14473 Potsdam,
Germany} 
\and 
K. P. Tod\footnote{
University of Oxford,
Mathematical Institute,
24-29 St. Giles,
Oxford OX1 3LB,
UK}
}

\maketitle

\begin{abstract}
The dynamics of a class of cosmological models with collisionless matter
and four Killing vectors is studied in detail and compared with that of
corresponding perfect fluid models. In many cases it is possible to 
identify asymptotic states of the spacetimes near the singularity or
in a phase of unlimited expansion. Bianchi type II models show oscillatory
behaviour near the initial singularity which is, however, simpler than
that of the mixmaster model.
\end{abstract}

\section{Introduction}

In studies of the dynamics of spatially homogeneous cosmological models
it is usual to choose a perfect fluid with linear equation of state to
describe the matter. The book \cite{wainwright97} provides an excellent 
guide to the
subject. In view of the fact that this restriction is made so 
frequently in the literature, it is natural to pose the question to what 
extent the conclusions obtained would change if the matter model were chosen 
differently. In \cite{rendall96} it was shown that in the case of 
collisionless matter 
described by the Vlasov equation significant changes can occur in comparison 
with the case of a perfect fluid. More specifically, it was shown that a 
solution of Bianchi type I exists whose qualitative behaviour near the 
initial singularity is different from that of any spacetime of that Bianchi
type whose matter content is a fluid with a physically reasonable equation of 
state, linear or nonlinear. In the following this analysis will be 
generalized to show just how different models with collisionless matter
can be from models with perfect fluid having the same symmetry. Differences
are found in models of Bianchi type II (Theorem 4.2), Bianchi type III
(Theorem 5.2) and Kantowski-Sachs models (Theorem 5.1). These concern both
the initial singularity and phases of unlimited expansion. Perhaps the most 
striking case is that of the initial singularity in the Bianchi type II
models, where we find persistent oscillatory behaviour near the singularity.
This is quite different from the known behaviour of type II perfect fluid
models.

Our results will also illuminate another matter. In \cite{lukash74} Lukash 
and 
Starobinski gave a heuristic analysis of a locally rotationally symmetric
(LRS) model of Bianchi type I with collisionless matter consisting of
massless particles. Their conclusion was that in the expanding direction 
the model would isotropize so that at large times it would look like a
Friedman-Robertson-Walker model. On the one hand we are able to prove
rigorously that the heuristic analysis of \cite{lukash74} gives the correct 
result.
On the other hand we show that this result depends essentially on the
assumption of a symmetry of Bianchi type I. If this symmetry type is 
replaced by Bianchi type II (keeping the LRS assumption and massless
collisionless particles) then the anisotropy tends to a constant non-zero
value at large times.

The cosmological models studied in this paper are LRS spatially homogeneous 
spacetimes with matter described by the Vlasov equation for massless 
particles. The reason for imposing the LRS condition is that it allows the 
Vlasov equation to be solved explicitly so that the Einstein-Vlasov equations
reduce to a system of ordinary differential equations, albeit with
coefficients which are not explicitly known and depend on the chosen
initial data. The reason for choosing the particles to be massless is
that this allows a reduction of the system of ODE similar to that 
carried out for perfect fluids with a linear equation of state  by
Wainwright and Hsu \cite{wainwright89}. It has not proved possible to analyse 
the global
behaviour of solutions to our system of ODE completely. However a number
of partial results have been obtained which show that there is considerable
variety in the asymptotic behaviour of solutions near an initial 
singularity or during a phase of unlimited expansion. In particular,
the reflection symmetric LRS Bianchi type I solutions with massless particles 
are analysed completely with respect to their asymptotic behaviour, thus 
improving markedly on the results obtained on that class of spacetimes in  
\cite{rendall96}. 

The matter model used in the following will now be described. The matter
consists of particles of zero rest mass which propagate through 
spacetime freely without collisions. Each particle is affected by the
others only by the gravitational field which they generate collectively.
The wordline of each particle is a null geodesic. Each geodesic has a
natural lift to the tangent bundle of spacetime. Thus the geodesic equation
defines a flow on the tangent bundle. By means of the metric this may if
desired be transported to the cotangent bundle and here it will be convenient
to do so. The subset of the cotangent bundle consisting of all covectors
obtained by lowering the index of future-pointing null vectors,
which will be denoted by $P$, is invariant under the flow and thus the flow 
may be restricted to it. The basic matter field used to describe the 
collisionless particles is a non-negative real-valued function $f$ on $P$ 
which represents the density of particles with given position and momentum 
at a given time. Choosing appropriate coordinates $x^\alpha$ on spacetime 
and letting $(x^\alpha,p_\alpha)$ be the corresponding coordinates on the 
cotangent bundle, the manifold $P$ can be coordinatized by $(x^\alpha,p_a)$.
Here the convention is used that Greek and Roman indices run from $0$ to
$3$ and $1$ to $3$ respectively. We write $t$ for $x^0$ and it is assumed
that $t$ increases towards the future. The field equation for $f$, the Vlasov
equation, says geometrically that $f$ is constant along the geodesic flow.
In the coordinates just introduced its explicit form is:
\begin{equation}\label{vlasov}
\d f/\d t+(p^a/p^0)\d f/\d x^a+(\Gamma^\alpha_{b\gamma}
p_\alpha p^\gamma/p^0)\d f/\d p_b=0
\end{equation}
where $p^0$ is to be determined from $p^a$ by the relation 
$g_{\alpha\beta}p^\alpha p^\beta=0$ and indices are raised and lowered
using the spacetime metric $g_{\alpha\beta}$ and its inverse. In order to 
couple the Vlasov equation to the Einstein equation, it is necessary to 
define the energy-momentum tensor. It is given by
\begin{equation}\label{energymomentum}
T_{\alpha\beta}=-\int fp_\alpha p_\beta |g|^{-1/2}/p_0 dp_1 dp_2 dp_3
\end{equation}
In fact for Bianchi models it is more useful to replace the coordinate
components of the momentum used in these equations by components in a 
suitable frame. The only change in the equations is that the Christoffel 
symbols in the Vlasov equation are replaced by the connection coefficients 
in the given frame. For more information about the Vlasov equation in general 
relativity the reader is referred to \cite{ehlers73} and \cite{rendall97a}. 

Spatially homogeneous spacetimes fall into three broad classes, known as
Bianchi class A, Bianchi class B and Kantowski-Sachs 
(see \cite{wainwright97}). Each of
the two Bianchi classes can be further divided into Bianchi types. A
spatially homogeneous spacetime in one of the Bianchi classes
is called locally rotationally symmetric
if it has, in addition to the three Killing vector fields needed for 
spatial homogeneity, a fourth one. This can only happen for certain 
symmetry types. In class A the Bianchi types which allow an LRS
special case are I, II, VII${}_0$, VIII and IX. In class B it is types
III, V and VII${}_h$ which allow this \cite{maartens85}. The Kantowski-Sachs 
spacetimes 
automatically have a fourth Killing vector. There exist solutions of the
Einstein-Vlasov equations with $k=-1$ Robertson-Walker symmetry and these
have, in particular, Bianchi type V and Bianchi type VII${}_h$ symmetry with
any non-zero $h$. We did not
attempt to ascertain whether there are other examples of solutions of these
Bianchi types with LRS symmetry, and these types are not considered further
in this paper. A spatially homogeneous solution 
of the Einstein-Vlasov equations has by definition the property that both the
geometry and the phase space density of particles are invariant under
the group action defining the symmetry type. A similar remark applies
to an additional LRS symmetry. It would be nice if the invariance of
$f$ under the group in a Bianchi model could be expressed by the condition
that $f$ depends only on time and momentum
when expressed with respect to a left-invariant frame on the group defining 
the symmetry. Unfortunately, as
discussed in \cite{maartens90}, this does not work in general. It does work 
for all
LRS Bianchi models of class A and type III and for Kantowski-Sachs models
\cite{maartens85}. This is the reason why LRS models are relatively tractable. 
In the 
following we consider LRS models which are of Kantowski-Sachs type, or 
of Bianchi type I, II, III, VII${}_0$, VIII or IX.

In the next section it is shown how in the class of spacetimes of 
interest the Einstein-Vlasov equations with given initial data can be 
reduced to a system of ordinary differential equations. In fact two
systems are needed. The first includes the solutions of types I, II, VII${}_0$,
VIII and IX while the second includes those of types I and III and the
Kantowski-Sachs models. Note that the solutions of type I are represented
in both systems and understanding the Bianchi I case is central to 
analysing the general case. The analysis of the Bianchi I system is 
carried out in the third section. This is then used in Sections 4 and 5 to 
obtain results on the first and second systems of ODE respectively. In
the last section the results are summarized and their wider significance
is examined. An appendix collects together some results from the theory of 
dynamical systems used in the body of the paper.

\section{Reduction to an ODE problem}

In a spacetime with Bianchi symmetry the metric can be written in the
form
\begin{equation}\label{bianchi}
ds^2=-dt^2+g_{ab}(t)\theta^a\otimes \theta^b
\end{equation}
where $\{\theta^a\}$ is a left-invariant coframe on the Lie group $G$ which 
defines the symmetry. The particular Bianchi type is determined by the 
structure constants of the Lie algebra of $G$. The extra symmetry which is 
present in the LRS case implies that the metric $g_{ab}(t)$ is diagonal, with 
two of the diagonal elements being equal \cite{maartens85}. Thus 
(\ref{bianchi}) simplifies to
\begin{equation}\label{lrs}
ds^2=-dt^2+a^2(t)(\theta^1)^2+b^2(t)((\theta^2)^2+(\theta^3)^2)
\end{equation}
for two functions $a(t)$ and $b(t)$ of one variable. If $k^\alpha$ is any 
Killing vector field then the function $p_\alpha k^\alpha$ on the cotangent
bundle is constant along geodesics and hence satisfies the Vlasov equation.
Any function of quantities of this type for different Killing vectors also
satisfies the Vlasov equation. The Killing vectors on a spacetime with 
Bianchi symmetry include those defined by right-invariant vector fields on 
the Lie  group $G$ but the result of evaluating a left-invariant one-form on 
one of these is not in general constant. Thus we cannot simply solve the 
Vlasov equation by choosing an arbitrary function of the components $p_a$ with
respect to a left-invariant basis. However for the LRS spacetimes of Bianchi
class A or type III considered here a function of the form 
$f(t,p_1,p_2,p_3)=f_0(p_1,p_2^2+p_3^2)$
does satisfy the Vlasov equation and in fact is the most general solution
with the full LRS symmetry \cite{maartens85}. Here $p_1$, $p_2$ and $p_3$ are 
the components 
of the momentum in the coframe $\{\theta^a\}$. Since $f$ does not depend 
explicitly on time in this representation, the function $f_0$ can be 
identified with the initial datum for the solution of the Vlasov equation at 
a fixed time. A similar statement holds for Kantowski-Sachs spacetimes. The 
metric can be written in the form (\ref{lrs}) where $\theta^1$ is invariant 
under 
the symmetry group and $\theta^2$ and $\theta^3$ make up any (locally defined)
orthonormal coframe on the two-sphere. The expression $p_2^2+p_3^2$ is not 
changed by a change in orthonormal coframe and so it makes sense to consider 
the above form of $f$ in terms of $f_0$ in Kantowski-Sachs spacetimes as well.
If $f$ is of this form it satisfies the Vlasov equation. Thus the Vlasov 
equation has been solved explicitly in the class of spacetimes to be studied. 
It remains to determine the form of the Einstein equations. In fact, one 
further restriction will be imposed. The distribution function given above
is automatically an even function of $p_2$ and $p_3$. However it need not be
even in $p_1$. If it is even in $p_1$ we say, as in \cite{rendall96}, that 
the solution
is reflection symmetric. Only reflection symmetric solutions will be 
considered in the following. For convenience we say that a function of
$p_1$, $p_2$ and $p_3$ which depends only on $p_1$ and $p_2^2+p_3^2$ and
which is even in $p_1$ has special form.

If the Einstein equations are split as usual into constraints and evolution
equations then it turns out that in this class of spacetimes the momentum
constraint is automatically satisfied. Only the Hamiltonian constraint and
the evolution equations are left. The former is an algebraic relation
between $a$, $b$ and their time derivatives $da/dt$, $db/dt$. The latter
provide ordinary differential equations for the evolution of $a$ and $b$
which are second order in time. It will be convenient to write these 
equations in terms of some alternative variables. Consider first the 
mean curvature of the homogeneous hypersurfaces:
\begin{equation}\label{meancurv}
{\rm tr}k=-[a^{-1}da/dt+2b^{-1}db/dt]
\end{equation}
A new time coordinate $\tau$ can be defined by $\tau(t)=-\int_{t_0}^t
{\rm tr}k(t) dt$ for some arbitrary fixed time $t_0$. In the following a dot 
over a quantity denotes its derivative with respect to $\tau$. Now define:
\begin{eqnarray}\label{dimensionless}
q&=&b/a,                                         \nonumber\\
N_1&=&-\epsilon_1 (a/b^2)({\rm tr k})^{-1},      \nonumber\\
N_2&=&-\epsilon_2 a^{-1}({\rm tr k})^{-1},                \\
\Sigma_+&=&-3(b^{-1}db/dt)({\rm tr k})^{-1}-1,   \nonumber\\
B&=&-b^{-1}({\rm tr} k)^{-1}\nonumber
\end{eqnarray}
where $\epsilon_1$ and $\epsilon_2$ will be $-1$, $0$ or $1$, depending on 
the symmetry type considered. The variables $N_1$, $N_2$ and $\Sigma_+$ are
closely related to the variables of the same names used by Wainwright and
Hsu \cite{wainwright89}. (Note that we adopt the conventions of 
\cite{wainwright89} rather than those of \cite{wainwright97}, which differ 
by a factor of three in some places.)

Two systems of ODE will now be considered, which between them are equivalent
to the evolution part of the Einstein-Vlasov equations for all the
relevant symmetry types. 

The first system is:
\begin{eqnarray}\label{bianchiA}
\dot q&=&\Sigma_+ q                             \nonumber\\
\dot N_1&=&[-\f{1}{4} N_1(N_1-4N_2)
+\f{1}{3} (1-4\Sigma_++\Sigma_+^2)]N_1
\nonumber       \\
\dot N_2&=&[-\f{1}{4} N_1(N_1-4N_2)+\f{1}{3} 
(1+2\Sigma_++\Sigma_+^2)]N_2
       \\
\dot \Sigma_+&=&\f{3}{2} \{\f{1}{2} N_1^2
+\f{1}{6} N_1(N_1-4N_2)(1-2\Sigma_+)            \nonumber\\
&+&[-\f{1}{4} N_1(N_1-4N_2)
+\f{1}{3}(1-\Sigma_+^2)][\f{1}{3}(1-2\Sigma_+)-Q]\}\nonumber
\end{eqnarray}
Here $Q$ is defined to be 
\begin{equation}\label{Q}
Q(q)=q^2\left[{
\int f_0(p_i)p_1^2(q^2 p_1^2+p_2^2+p_3^2)^{-1/2} dp_1dp_2dp_3}
\over 
\int f_0(p_i)(q^2 p_1^2+p_2^2+p_3^2)^{1/2} dp_1dp_2dp_3
\right]
\end{equation}
where $f_0$ is a fixed smooth function of special form and compactly 
supported on $\R^3$. The Hamiltonian constraint is
\begin{equation}\label{hamiltonian}
16\pi\rho/(\tr k)^2=-\f{1}{2}N_1(N_1-4N_2)+\f{2}{3} (1-\Sigma_+^2)
\end{equation}
where $\rho$ is the energy density and to take account of the positivity of 
$\rho$, only the region satisfying the inequality
\begin{equation}\label{physical}
-\f{1}{2}N_1(N_1-4N_2)+\f{2}{3} (1-\Sigma_+^2)\ge 0
\end{equation}
is considered. Define submanifolds of this region by the following 
conditions: 
\begin{eqnarray*}
&&S_1:\ \ \ N_1=N_2=0                               \\
&&S_2:\ \ \ N_1\ne 0, N_2=0                         \\
&&S_3:\ \ \ N_1=0, N_2\ne 0                         \\
&&S_4:\ \ \ N_1\ne 0, N_2\ne 0, N_2=-q^2 N_1        \\
&&S_5:\ \ \ N_1\ne 0, N_2\ne 0, N_2=q^2 N_1
\end{eqnarray*}
The submanifolds $S_1$, $S_2$, $S_3$, $S_4$ and $S_5$
correspond to Bianchi types I, II, VII${}_0$, VIII and IX respectively.
To make the correspondence with spacetime quantities in these different
cases $(\epsilon_1,\epsilon_2)$ should be chosen to be $(0,0)$, $(1,0)$, 
$(0,1)$, $(-1,1)$ and $(1,1)$ respectively. Note that if $q$ is replaced by 
$\tilde q=q^{-1}$ in (\ref{bianchiA}) an almost identical system is obtained, 
with the 
sign in the first equation being reversed. 

The second system is:
\begin{eqnarray}\label{surfacesymm}
\dot q&=&\Sigma_+ q                                            \nonumber\\
\dot B&=&[\epsilon B^2+\f{1}{4}+\f{1}{12}(1-2\Sigma_+)^2]B              \\
\dot\Sigma_+&=&\f{3}{2}\{-\f{2}{3}\epsilon B^2 (1-2\Sigma_+)
+[\epsilon B^2+\f{1}{3} 
(1-\Sigma_+^2)][\f{1}{3}(1-2\Sigma_+)-Q]\}\nonumber
\end{eqnarray}
where $\epsilon$ belongs to the set $\{-1,0,1\}$. Only the region satisfying
the inequality
\begin{equation}\label{physicalsurf}
2\epsilon B^2+\f{2}{3} (1-\Sigma_+^2)\ge 0
\end{equation}
is considered. The cases $\epsilon=-1$, $\epsilon=0$ and $\epsilon=1$ 
correspond to Bianchi type III, Bianchi type I and Kantowski-Sachs 
respectively. Note that the restriction of the system (\ref{bianchiA}) to 
$S_1$ is 
identical to the system consisting of the first and third equations of 
(\ref{surfacesymm})
for $\epsilon=0$. This restricted system
will be referred to in the following as the Bianchi I system. It was
introduced in section 6 of \cite{rendall96} with slightly different variables.

If a solution of (\ref{bianchiA}) and a fixed $f_0$ are given, it is possible 
to
construct a spacetime as follows. Suppose that $\tau=0$ is contained in the 
domain of definition of the solution. Since the system is autonomous this is
no essential restriction. Choose a negative number $H_0$. Define 
\begin{equation}\label{density}
\rho=(1/16\pi)H_0^2[-\f{1}{2}N_1(0)(N_1(0)-4N_2(0))
+\f{2}{3} (1-\Sigma_+^2(0))]
\end{equation}
Let
\begin{equation}\label{densityint}
I=\int f_0(p_i)[(q(0))^2 p_1^2+p_2^2+p_3^2]^{1/2}dp_1 dp_2 dp_3
\end{equation}
and define
\begin{eqnarray}\label{scalefactors}
a_0&=&\rho^{-1/4}I^{1/4}(q(0))^{-3/4} \nonumber\\
b_0&=&\rho^{-1/4}I^{1/4}(q(0))^{1/4}
\end{eqnarray}
In terms of these quantities we can define an initial metric by
\begin{equation}\label{initialmetric}
a_0^2(\theta^1)^2+b_0^2((\theta^2)^2+(\theta^3)^2)
\end{equation}
Similarly, we can define an initial second fundamental form by
\begin{equation}\label{sff}
-\f{1}{3}(1-2\Sigma_+(0))H_0a_0^2(\theta^1)^2
+\f{1}{3}(1+\Sigma_+(0))H_0b_0^2((\theta^2)^2+(\theta^3)^2)
\end{equation}
These data satisfy the constraints by construction.
Consider now the spacetime which evolves from these initial data. It is of the 
form (\ref{bianchi}). For the Einstein-Vlasov system in a spacetime of the 
form (\ref{bianchi})
with a fixed time-independent distribution function is a system of second
order ODE which has solutions corresponding to data for $(a,b,da/dt,db/dt)$.
These data can be chosen so as to reproduce the data of interest for the
Einstein-Vlasov system by choosing $da/dt=\f{1}{3}(1-2\Sigma_+(0))H_0a$ and
$db/dt=-\f{1}{3}(1+\Sigma_+(0))H_0b$ for $t=t_0$. This spacetime defines a 
solution
of (\ref{bianchiA}) via (\ref{dimensionless}). (Note that $t=t_0$ corresponds 
to $\tau=0$.)
Thus the two solutions are identical. In this way a spacetime
has been constructed which gives rise to the solution of (\ref{bianchiA}) we 
started with.
This spacetime may be obtained more explicitly if desired. In order to do 
this, first solve the equation:
\begin{equation}\label{meanevolution}
\d_\tau (\tr k)=-[-\f{1}{4}N_1(N_1-4N_2)+\f{1}{3}
(2+\Sigma_+^2)]\tr k
\end{equation}
with initial data $H_0$. Then $\rho$ can be obtained from the
Hamiltonian constraint (\ref{hamiltonian}). The definition of $\rho$ in terms 
of $f_0$ can 
then be combined with $q$ to give $a$ and $b$ as in (\ref{scalefactors}). 
Finally $t$ can be
obtained from $\tr k$. All the considerations here in the case of 
(\ref{bianchiA}) are
equally applicable in the case of (\ref{surfacesymm}). The analogue of 
equation (\ref{meanevolution}) is
\begin{equation}\label{meanevolutionsurf}
\d_\tau (\tr k)=-[\epsilon B^2+\f{1}{3}(2+\Sigma_+^2)]\tr k
\end{equation}

Solutions of the Einstein equations with matter described by a perfect fluid 
with a linear equation of state $p=(\gamma-1)\rho$ which belong to one of the 
symmetry types studied in the case of collisionless matter in the following 
can be described by equations very similar to (\ref{bianchiA}) and 
(\ref{surfacesymm}). The similarity
is particularly great in the case $\gamma=\f{4}{3}$ (radiation fluid). In 
that 
case the only difference is that the function $Q(q)$ should be replaced by 
the constant value $\f{1}{3}$. This leads to a decoupling of the first 
equation in each system, so that it is possible to restrict attention to the 
remaining equations when investigating the dynamics. (This last remark also 
applies to the system obtained for other values of $\gamma$.) The equation for 
$q$ can be integrated afterwards if desired.

In \cite{rendall96} it was proved that $Q(q)$ as defined in (\ref{Q}) tends to 
zero as $q$ 
tends to zero and that if $Q(0)$ is defined to be zero the resulting 
extension of $Q$ is $C^1$ with $Q'(0)=0$. This means in particular that the 
dynamical system (\ref{bianchiA}) has a well-defined $C^1$ extension to $q=0$. 
In a 
similar way it can be shown that if a function $\tilde Q$ is defined by 
$\tilde Q(\tilde q)=Q(q)$ then $\tilde Q$ can be extended in a $C^1$ manner 
to $\tilde q=0$ in such a way that $\tilde Q(0)=1$ and $\tilde Q'(0)=0$. For
$1-\tilde Q=(\rho-T^1_1)/\rho=2T^2_2/\rho$ and this last expression is
$O(\tilde q^{4/3})$ as $\tilde q\to 0$ by Lemma 4.2 of \cite{rendall96}. By 
using a coordinate $\hat q=q/(q+1)$ it is possible to map the system 
(\ref{bianchiA}) 
with $q$ ranging from zero to infinity onto a region with $\hat q$ ranging
from zero to one. Moreover the system extends in a $C^1$ manner to the
boundary components $\hat q=0$ and $\hat q=1$. The coordinate $\hat q$
has been introduced purely to demonstrate that the system (\ref{bianchiA}) 
can be
smoothly compactified in the $q$-direction. For computations it is more 
practical to use the local coordinates $q$ and $\tilde q$. In particular,
these considerations allow us to regard the Bianchi I system as being defined
on a compact set. 

The compactification of the system (\ref{bianchiA}) is defined on a region 
with boundary.
Different parts of the boundary are given by $\hat q=0$, $\hat q=1$ and the
case of equality in (\ref{physical}). The complement of the boundary will be 
called
the interior region in the following. A solution which lies in the interior
region corresponds to a smooth non-vacuum solution of the Einstein-Vlasov 
equations. A solution which lies in the part of the boundary where 
(\ref{physical})
becomes an equality corresponds to a solution of the vacuum Einstein 
equations. A solution which lies in the part of the boundary given by 
$\hat q=0$ or $\hat q=1$ corresponds to a distributional solution of the 
Einstein-Vlasov equations, as will be explained in more detail below. The
system (\ref{surfacesymm}) can be compactified in a way very similar to that 
taken in the case of (\ref{bianchiA}). The comments on the interpretation of 
different types of solutions of the compactification of (\ref{bianchiA}) just 
made also apply to the compactification of (\ref{surfacesymm}), with 
(\ref{physical}) being replaced by (\ref{physicalsurf}).

Consider now the stationary points of the system (\ref{bianchiA}), or rather 
of its
compactification. (This distinction will not always be made explicitly in
what follows.) In section 4 it will be shown that all stationary points where 
$q$ has a finite non-zero value belong to the subset $S_1$ corresponding to 
solutions of type I. In particular, they correspond to stationary points of 
the Bianchi I system, which will be studied in detail in the next section.

\section{The Bianchi I system}

It turns out that the Bianchi I system plays a central role in the dynamics 
of solutions of the systems (\ref{bianchiA}) and (\ref{surfacesymm}). In this 
section the asymptotic
behaviour of solutions of this system is determined, both for $\tau\to-\infty$
(approach to the singularity) and for $\tau\to\infty$ (unlimited expansion).

The first step in analysing the Bianchi I system is to determine the 
stationary points. This can be done using the fact, proved in 
\cite{rendall96}, that
$Q$ is strictly monotone for $q>0$ so that there is a unique $q_0$ with
$Q(q_0)=\f{1}{3}$. With this information it is straightforward to show that
the coordinates of the stationary points in the $(q,\Sigma_+)$ plane are 
$(q_0,0)$, $(0,-1)$, $(0,\f{1}{2})$, $(0,1)$, $(\infty,-1)$ and 
$(\infty,1)$. 
Here $q=\infty$ is to be interpreted as $\tilde q=0$ or $\hat q=1$. Call 
these points $P_1,\ldots,P_6$ respectively (see figure 1).
\begin{figure}
\begin{center}
\includegraphics[width=8cm,height=5cm,angle=270]{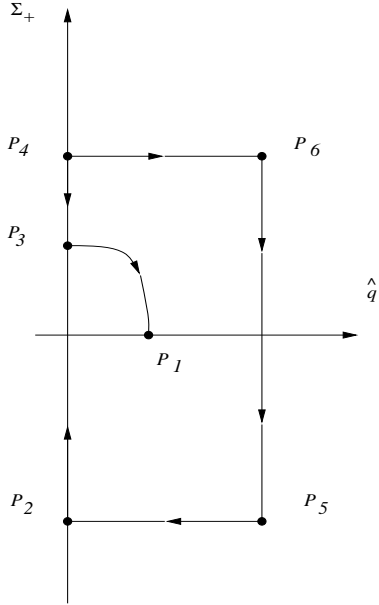}
\end{center}
\label{Figure 1}
\caption{The $(\hat{q},\Sigma_+)$ plane and the fixed points for Bianchi
type I.}
\end{figure}
The next step is to linearize the system about the stationary points. Recall 
that a stationary point is called hyperbolic if none of its eigenvalues are 
purely imaginary. In the following we call a stationary point degenerate if it 
is not hyperbolic. The point $P_1$ is a hyperbolic sink while $P_4$ is a 
hyperbolic source. The points $P_2$, $P_3$ and $P_6$ are hyperbolic saddles 
while $P_5$ is degenerate, with one zero eigenvalue. 

Before proceeding further, we state the main result of this section.

\vskip 10pt\noindent
{\bf Theorem 3.1} If a smooth non-vacuum reflection-symmetric LRS solution of 
Bianchi type I of the Einstein-Vlasov equations for massless particles is 
represented as a solution of (\ref{bianchiA}) with $N_1=N_2=0$ then for 
$\tau\to\infty$ 
it converges to the point $P_1$. For $\tau\to -\infty$ either

\noindent
(i) it converges to $P_1$ and in that case it stays for all time at the point 
$P_1$ or
\next
(ii) it converges to the point $P_3$ and it belongs to the unstable manifold 
of $P_3$ or
\next
(iii) it converges to $P_4$
\next
All of these cases occur, and (iii) is the generic case in the sense that it
occurs for an open dense set of initial data. 

This will be proved in a series of lemmas. Terminology from the theory of
dynamical systems which may be unfamiliar to the reader is explained in 
the appendix.

\noindent
{\bf Lemma 3.1} If a solution of the Bianchi I system in the interior
enters the region $\Sigma_+>1/2$ then for $\tau\to -\infty$ it belongs to 
case (iii) of Theorem 3.1. A solution of the Bianchi I system in the interior 
has no $\omega$-limit points with $\Sigma_+\ge 1/2$. 

\noindent
{\bf Proof} A solution of the Bianchi I system satisfies $\dot\Sigma\le
-\f{1}{2}(1-\Sigma_+^2)Q$ 
when $\Sigma_+>1/2$ and so for any solution which enters the given region,
$\Sigma_+$ is nondecreasing towards the past and it is in the region for all
earlier times. If $\Sigma_+$ did not tend to $1$ as $\tau\to -\infty$ then
we would have $\dot\Sigma_+\le -C<0$ at early times, a contradiction. Once we
know that $\Sigma_+\to 1$ as $\tau\to -\infty$ it follows immediately that
$q\to 0$. Thus the solution converges to $P_4$. Consider now the forward 
time direction. Since $\Sigma_+$ is positive, $q$ is increasing. This means
that $Q$ is increasing. The inequality $\dot\Sigma\le-CQ$ for a constant 
$C>0$ then shows that the solution must leave the region of interest in 
finite time, so that there can be no $\omega$-limit point with 
$\Sigma_+\ge 1/2$.

\vskip 10pt\noindent
For convenience an interior  solution which does not tend to $P_4$ as 
$\tau\to -\infty$ will be called exceptional. Thus Lemma 3.1 says that an 
exceptional solution cannot intersect the region $\Sigma_+>1/2$.

\noindent
{\bf Lemma 3.2} The $\alpha$-limit set of an exceptional solution 
cannot intersect the boundary at any point except $P_3$. If it does 
intersect the boundary at $P_3$ it belongs to case (ii) of Theorem 3.1
as $\tau\to -\infty$. The $\omega$-limit set of any interior solution cannot 
intersect the boundary at all. 

\noindent
{\bf Proof} Let $(q,\Sigma_+)$ be a point of the $\alpha$-limit set of an
exceptional solution which lies on the boundary. If $q=\infty$ then the whole 
orbit passing through that point belongs to the $\alpha$-limit set. This 
implies that the solution must intersect the region $\Sigma_+> 1/2$, a
contradiction.  Thus in fact $q<\infty$. If $\Sigma_+=-1$ then all points
with  $\Sigma_+=-1$ must be in the $\alpha$-limit set, in particular $P_5$.
By Lemma A2 of the appendix, it follows that a point of the centre manifold 
of $P_5$ lies in the $\alpha$-limit set. However, this centre manifold is
given by $q=\infty$ and so we again obtain a contradiction. Hence 
$\Sigma_+>-1$. If $q=0$ and $\Sigma_+<1/2$ then all points satisfying these
conditions must be in the $\alpha$-limit set, in particular $P_2$. But then
an application of Lemma A1 of the appendix leads to a contradiction. Thus no
point on the boundary other than $P_3$ is possible. A further application of
Lemma A1 shows that in this case the solution must lie on the unstable 
manifold of $P_3$. In a similar way it is
possible to show that if any point of the boundary belonged to the 
$\omega$-limit set of an interior solution then some point with 
$\Sigma_+>1/2$ would do so. However we know from Lemma 3.1 that this is
impossible. 

\noindent
{\bf Proof of Theorem 3.1} First the Poincar\'e-Bendixson theorem will be
applied to the restriction of the Bianchi I system to the interior with
the point $P_1$ removed. In general the $\alpha$- and $\omega$-limit sets of 
an orbit of a dynamical system can be very complicated, but in two dimensions
(and the Bianchi I system is two-dimensional) things are a lot simpler.
Complicated situations are still possible and these play an important role
in Hilbert's sixteenth problem (see e.g. \cite{arnold88}, p. 104). However 
many 
pathologies are ruled out by the Poincar\'e-Bendixson theorem, which is 
stated in the Appendix (Theorem A2).

Given an interior solution, suppose that $P_1$ does not belong
to the $\omega$-limit set. By Lemma 3.2 no point of the boundary belongs to
the $\omega$-limit set either. Since $P_1$ is a hyperbolic sink 
it follows that there must be a neighbourhood of $P_1$ which does not 
intersect the $\omega$-limit set. Thus the solution remains in a compact set
of the interior with the point $P_1$ removed as $\tau\to\infty$. Then Theorem 
A2 implies the existence of a non-stationary periodic orbit of the Bianchi I 
system. In fact the existence of periodic solutions of the Bianchi I system 
can be ruled out by the presence of a Dulac function. (For a discussion of 
this concept see \cite{wainwright97}.) Define a function $F(q)$ by 
\begin{equation}\label{primitive}
F(q)=\int f_0(p_i)(q^2 p_1^2+p_2^2+p_3^2)^{1/2} dp_1dp_2dp_3
\end{equation}
Then $Q=(q/F)F'$. For $q>0$ and $|\Sigma_+|<1$ let
\begin{equation}\label{dulac}
G(q,\Sigma_+)=q^{-1}F^{1/2}(1-\Sigma_+^2)^{-3/2}
\end{equation}
and denote the vector field defining the Bianchi I system by $X$. Then
${\rm div}(GX)$ is negative. In fact it is a constant multiple of 
$q^{-1}F^{1/2}(1-\Sigma_+^2)^{-1/2}(2-\Sigma_+)$. This means that 
$G$ is a Dulac function. It follows that the Bianchi I system has no periodic 
solutions. It can be concluded that $P_1$ does lie in the $\omega$-limit set.
But since $P_1$ is a hyperbolic sink, this implies, via the Hartman-Grobman
theorem (cf. Theorem A1), that the $\omega$-limit set consists of $P_1$ 
alone, which proves the first part of the theorem.

To prove the remainder of the theorem we can assume without loss of generality
that the solution is exceptional and that it does not lie on the unstable 
manifold of $P_3$. If $P_1$ were not in the $\alpha$-limit set then we
would get a contradiction by the Poincar\'e-Bendixson theorem and the
absence of periodic orbits. Hence $P_1$ must belong to the $\alpha$-limit
set, and since $P_1$ is a hyperbolic sink, the only possibility left is
case (i) of the theorem.

\vskip 10pt\noindent
The conclusion of this theorem can be summarized in words as follows. All
solutions isotropize in the expanding direction. The initial singularity
is generically a cigar singularity but there are exceptional cases where
it is a barrel or point singularity. (For this terminology see 
\cite{wainwright97}, p. 30.)
Note for comparison that if the Vlasov equation is replaced by the Euler 
equation for a fluid satisfying a physically reasonable equation of state
then there are no barrel singularities and all solutions which are not
isotropic have cigar or pancake singularities (see \cite{rendall96}). The
pancake singularities are as common as the cigar singularities. In
particular this means that for fluid solutions of Bianchi type I cigar
singularities are {\it not} generic. All solutions 
isotropize in the expanding direction. Note that the \lq reasonable\rq\ 
equations of state include those of the form $p=k\rho$ with $0\le k<1$. The 
solutions of the Einstein-Vlasov equations approach an isotropic fluid 
solution with equation of state $p=\f{1}{3}\rho$ in the sense that the 
tracefree
part of the spatial projection $T_{ij}$ of the energy-momentum tensor divided 
by the energy density $\rho$ approaches zero, while $\tr T/\rho=1$. The 
latter relation is always true for kinetic theory with massless particles
and for a radiation fluid (equation of state $p=\f{1}{3}\rho$).

\section{Other class A models}

This section is concerned with the models of class A, as described by the 
system (\ref{bianchiA}). Only limited statements will be made about types VIII 
and IX.
Even in the a priori simpler case of a perfect fluid with linear equation of 
state it is difficult to analyse LRS models of type VIII and IX. (For 
information on what is known about that case, see 
\cite{uggla90}, \cite{uggla91} and section 8.5
of \cite{wainwright97}). A major difficulty is that in these cases
the domain of definition of the dynamical system is non-compact. This 
allows the possibility that there may be anomalous solutions similar to
those encountered in \cite{rendall97b}. Type I was analysed in the previous 
section
and we will see that the analysis of type VII${}_0$ can be reduced to that 
case in a relatively straightforward way. The most interesting results are
obtained for Bianchi type II.

We start with a theorem on Bianchi type VII${}_0$, which is a close analogue 
of Theorem 3.1.

\vskip 10pt\noindent
{\bf Theorem 4.1} If a smooth non-vacuum reflection-symmetric LRS solution of 
Bianchi type VII${}_0$ of the Einstein-Vlasov equations for massless particles
is represented as a solution of (\ref{bianchiA}) with $N_1=0$ then for 
$\tau\to\infty$ 
the pair $(q,\Sigma_+)$ converges to $(q_0,0)$ while $N_2$ increases without
limit. $N_2$ tends to zero as $\tau\to -\infty$ while the pair $(q,\Sigma_+)$ 
either

\noindent
(i) converges to $P_1$ and in that case it stays for all time at the point 
$P_1$ or
\next
(ii) converges to the point $P_3$ and belongs to the unstable manifold 
of $P_3$ or
\next
(iii) converges to $P_4$
\next
All of these cases occur, and (iii) is the generic case in the sense that it
occurs for an open dense set of initial data.

\noindent
{\bf Proof} When $N_1=0$ the third equation in (\ref{bianchiA}) becomes
\begin{equation}\label{ntwo}
\dot N_2=\f{1}{3} (1+\Sigma_+)^2 N_2
\end{equation}
while the equations for $\Sigma_+$ and $q$ do not involve $N_2$. The latter
equations form a subsystem which is identical to the equations for Bianchi
type I, so that the situation is again as in figure 1. The qualitative
behaviour of their solutions has been analysed in
Theorem 3.1. All that remains to be done is then to put that information 
into equation (\ref{ntwo}) and read off the behaviour of $N_2$. The 
expression $(1+\Sigma_+)^2$ is strictly positive for a non-vacuum solution
of type VII${}_0$, due to (\ref{physical}). Moreover it is bounded by four. 
Thus the
solution has the property that the sign of $N_2$ remains constant and the
solution exists globally in $\tau$. It is also clear that $N_2\to\infty$
as $\tau\to\infty$ and that $N_2\to 0$ as $\tau\to -\infty$.  

\vskip 10pt\noindent
There is a simple explanation for the close relation between the Bianchi
I and Bianchi VII${}_0$ solutions. They are in fact the same spacetimes 
parametrized in two different ways. The full four-dimensional isometry group 
has a subgroup of Bianchi type I and a one-parameter family of subgroups of 
Bianchi type VII${}_0$. 

Next we turn to the solutions of type II. It will be shown that the 
stationary points of (the compactification of) (\ref{bianchiA}) which lie in 
the closure 
of $S_2$ are the points $P_1,\ldots,P_6$ which we know already together with 
one additional point $P_7$, which has coordinates $(\infty,\f{2\sqrt2}{5},
\f{1}{5})$ (see figure 2). The corresponding distributional solution of the 
Einstein-Vlasov equations will be discussed in detail below. 
\begin{figure}
\begin{center}
\includegraphics[width=10cm,height=12cm,angle=270]{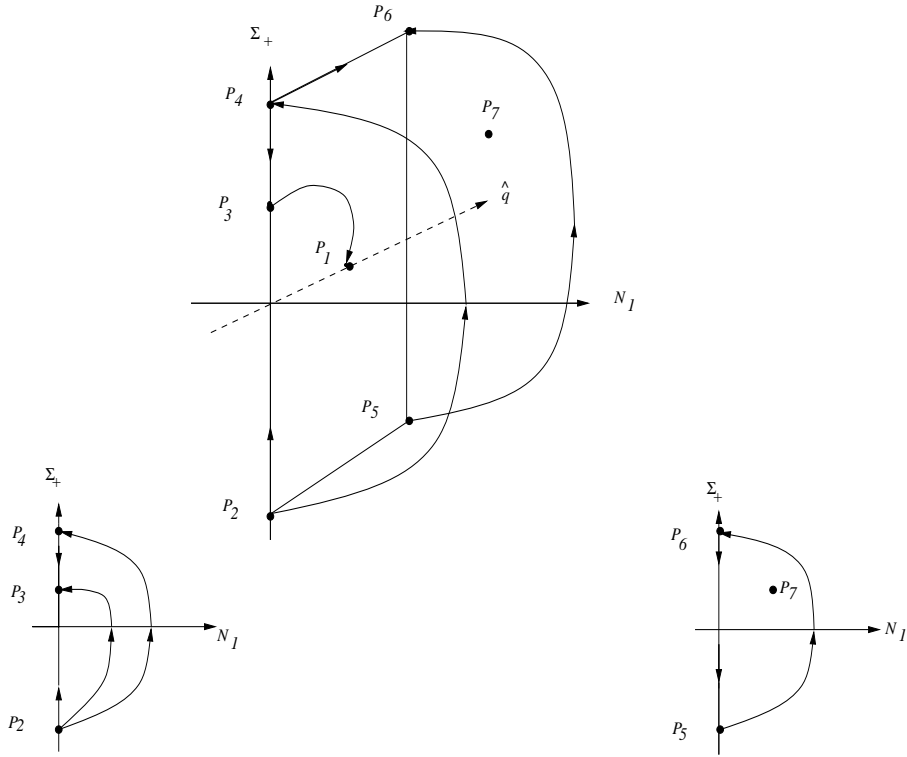}
\end{center}
\label{Figure 2}
\caption{The $(\hat{q},\Sigma_+,N_1)$ space and the fixed points for Bianchi
type II; the lower two figures show the phase portraits on the
end-faces $\hat{q}=0,1$}
\end{figure}
\vskip 10pt\noindent
{\bf Theorem 4.2} If a smooth non-vacuum reflection symmetric LRS solution of 
Bianchi type II of the Einstein-Vlasov equations for massless particles is 
represented as a solution of (\ref{bianchiA}) with $N_2=0$ then for 
$\tau\to\infty$ the 
solution converges to $P_7$. For $\tau\to -\infty$ either:

\noindent
(i) the solution converges to $P_1$ or
\next
(ii) the $\alpha$-limit set of the solution consists of the points $P_2$, 
$P_4$, $P_5$ and $P_6$ together with the orbits connecting $P_2$ to $P_5$,
$P_5$ to $P_6$ and $P_6$ to $P_4$ in the set $N_1=0$ and the stable manifold
of $P_4$ which connects $P_4$ with $P_2$ via the vacuum boundary. In 
particular $\liminf_{\tau\to-\infty}\Sigma_+=-1$, 
$\limsup_{\tau\to-\infty}\Sigma_+=+1$,
$\liminf_{\tau\to -\infty}q(\tau)=-\infty$ and
$\limsup_{\tau\to -\infty}q(\tau)=\infty$. 

\noindent
Both of these cases occur and (ii) is the generic case.

\vskip 10pt\noindent
This theorem shows that while models of Bianchi type II have simple behaviour
in the expanding phase, all tending to a single attractor, the behaviour
near the initial singularity is in general oscillatory, and quite different 
from the Bianchi type I case. Note also that the models of type II do not 
isotropize as $\tau\to\infty$, which is another important difference from the 
type I models. 

A first important step in proving Theorem 4.2 is to use the identity
\begin{equation}\label{liapunov}
\d/\d\tau (q^{4/3}N_1)=q^{4/3}N_1[-\f{1}{4}N_1(N_1-4N_2)
+\f{1}{3} (1-\Sigma_+^2)
+\f{2}{3}\Sigma_+^2]
\end{equation}
which holds for any solution of (\ref{bianchiA}). This is done in the 
following lemma.

\noindent
{\bf Lemma 4.1} For any solution in the interior of $S_2$ the following
statements hold. As $\tau\to\infty$ $q$ tends to $\infty$. Either the 
$\alpha$-limit set is contained in the set $q=0$ or it contains one of the 
points $P_1,\ldots,P_6$.

\noindent
{\bf Proof} For a non-vacuum solution strict inequality holds in 
(\ref{physical})  and 
hence $q^{4/3}N_1$ is strictly increasing where it is non-zero. This means 
that as long as $q$ is finite and the Bianchi type is II this quantity is 
always increasing. Since $S_2$ is compact, solutions exist globally in $\tau$.
As $\tau$ tends to plus or minus infinity the solution must go to the 
boundary of $S_2$. Equation (\ref{liapunov}) shows that if $q^{4/3}N_1$ tends 
to a 
finite non-zero limit in either time direction then $\Sigma_+^2$ is integrable
on a half-infinite time interval. The derivative of this quantity is
bounded and these two facts together imply that it must tend to zero
in the limit. The same argument applies to the quantity appearing in 
(\ref{physical})
and so it must also tend to zero in the limit. Under these conditions 
$N_1\to \f{2}{\sqrt 3}$ and $\dot\Sigma_+\to \f{4}{3}$, a contradiction.
It can be concluded that $\lim_{\tau\to\infty}(q^{4/3}N_1)(\tau)=\infty$ and
$\lim_{\tau\to -\infty}(q^{4/3}N_1)(\tau)=0$. {}From the first of these 
statements and the boundedness of $N_1$ it follows that $q$ tends to $\infty$
as $\tau$ tends to $\infty$. The fact that $q^{4/3}N_1$ tends to zero in the 
contracting direction implies that the $\alpha$-limit set is contained in the 
union of the sets $q=0$ and $N_1=0$. Suppose the $\alpha$-limit set contains 
some point for which $q\ne 0$. This must belong to the Bianchi I set. Thus 
the $\alpha$-limit set contains a solution of Bianchi type I. Using Theorem 
3.1, we conclude that the $\alpha$-limit set contains one of the points $P_1,
\ldots,P_6$.

\vskip 10pt
The next lemma gives information about the nature of the stationary points
on $S_2$. We already know from Lemma 4.1 that these stationary points can
only occur for $N_1=0$, $q=0$ or $q=\infty$. The stationary points $P_1,
\dots,P_6$ will be investigated first. The equations for $q=0$ and $q=\infty$
will be studied in detail later.

\noindent
{\bf Lemma 4.2} The stationary points $P_1,\ldots,P_4$ and $P_6$ of the 
restriction of the system (\ref{bianchiA}) to $S_2$ are hyperbolic saddles, 
while $P_5$ 
is degenerate. The stable manifold of $P_1$ is given by $N_1=0$.
The stable and unstable manifolds of $P_2$ are given by $\Sigma_+=-1$, $N_1=0$
and $q=0$ respectively. The stable manifold of $P_3$ is given by 
$q=0$. The stable and unstable manifolds of $P_4$ are given 
by $q=0$, $N_1^2=\f{4}{3}(1-\Sigma_+^2)$ and $N_1=0$ respectively. The 
stable and 
unstable manifolds of $P_6$ are given by $N_1^2=\f{4}{3}(1-\Sigma_+^2)$ and
and $q=\infty$, $N_1=0$ respectively. The unstable manifold of $P_5$ is given 
by $N_1^2=\f{4}{3}(1-\Sigma_+^2)$. The set $q=\infty$, $N_1=0$ is a centre 
manifold for $P_5$. 

\noindent
{\bf Proof} All that needs to be done is to compute the linearizations of
the system about the given points and to note that the manifolds named in 
the statement of the theorem are all invariant. Linearizing the restriction
of (\ref{bianchiA}) to $N_2=0$, and setting $N_1=0$ in the result, gives the 
system 
(a bar denotes a linearized quantity):
\begin{eqnarray}\label{linearized}
d\bar q/d\tau&=&\Sigma_+\bar q+q\bar\Sigma_+          \nonumber\\
d\bar N_1/d\tau&=&\f{1}{3}(1-4\Sigma_++\Sigma_+^2)\bar N_1     \\
d\bar \Sigma_+/d\tau&=&[-\f{1}{3}(1+\Sigma_+-3\Sigma_+^2)+\Sigma_+Q(q)]
\bar\Sigma_+
-\f{1}{3} Q'(q)(1-\Sigma_+^2)\bar q\nonumber
\end{eqnarray}
The linearization about $P_1$ has eigenvalues $\f{1}{3}$ and 
$-\f{1}{6}\pm\f{1}{2}\sqrt{\f{1}{9}
-\f{4}{3} q_0 Q'(q_0)}$. The invariant subspace of the 
linearization corresponding to the eigenvalues with negative real parts is 
the tangent space to $N_1=0$. The linearizations about $P_2$, $P_3$ and $P_4$ 
are diagonal with diagonal entries $(-1,2,1)$, $(\f{1}{2},
-\f{1}{4},-\f{1}{4})$ and 
$(1,-\f{2}{3},\f{1}{3})$ respectively. Since $P_5$ and $P_6$ lie at
$q=\infty$, we must change to the coordinate $\tilde q$ to study the 
linearizations at these points. They are diagonal with diagonal elements 
$(1,2,0)$ and $(-1,-\f{2}{3},\f{4}{3})$.

\vskip 10pt\noindent

Next the limiting systems for $q=0$ and $q=\infty$ will be examined.

\noindent
{\bf Lemma 4.3} Consider the restriction of the system (\ref{bianchiA}) to 
the set given by the equations $N_2=q=0$. For any solution which does not 
belong to the vacuum boundary and which does not satisfy $N_1=0$, the 
$\alpha$-limit set is the point $P_2$ and the $\omega$-limit set is the 
point $P_3$.

\noindent
{\bf Proof} First it will be shown that the solution cannot be stationary.
The equation for $\Sigma_+$ shows that $\dot\Sigma_+>0$ if 
$\Sigma_+<\f{1}{2}$.
Thus at a stationary point $\Sigma_+\ge\f{1}{2}$. On the other hand, the 
equation 
for $N_1$ shows that at a stationary point
\begin{equation}\label{sigmaplus}
(\Sigma_+-2)^2=3+N_1^2\ge 3
\end{equation}
Using the fact that $\Sigma_+\le 1$ it follows that $\Sigma_+\le 2-\sqrt 3
<\f{1}{2}$. Next, it follows from Theorem 3.1 on p. 150 of 
\cite{hartman82} that the 
solution cannot be periodic. It can be concluded using the 
Poincar\'e-Bendixson theorem (Theorem A2 of the appendix) that the $\alpha$- 
and $\omega$-limit sets are contained in the boundary of the region. The 
behaviour of solutions on the boundary is easily determined. The nature
of the stationary points on the boundary can be read off from Lemma 4.2.
$P_2$ is a hyperbolic source, $P_3$ is a hyperbolic sink and $P_4$ is a
hyperbolic saddle. The last fact means, using Lemma A1 of the appendix, 
that $P_4$ cannot be in the $\alpha$- or $\omega$-limit set unless $P_2$ or 
$P_3$ is also. Thus it can be concluded that the $\alpha$- and $\omega$-limit 
sets must contain either $P_2$ or $P_3$ and then the conclusion follows 
easily.

\vskip 10pt\noindent
The system given by $N_2=0$ and $q=\infty$ is more complicated.

\noindent
{\bf Lemma 4.4} Consider the restriction of the system (\ref{bianchiA}) to 
the set 
given by the equations $N_2=0$ and $q=\infty$. For any solution which does not 
belong to the vacuum boundary and which does not satisfy $N_1=0$, the 
$\alpha$-limit set consists of all points which are either on the vacuum
boundary or satisfy $N_1=0$. The $\omega$-limit set is the point $P_7$.

\noindent
{\bf Proof} Define a function $Z$ by
\begin{equation}\label{liapunovII}
Z=N_1^{1/2}[-\f{1}{2} N_1^2+\f{2}{3}(1-\Sigma_+^2)]^{3/4}
(1-\f{1}{5}\Sigma_+)^{-2}
\end{equation}
This function is well-defined and continuous on $S_2$ and smooth away from
$N_1=0$ and $-\f{1}{2} N_1^2+\f{2}{3}(1-\Sigma_+^2)=0$. Its 
derivative is given by
\begin{equation}\label{liapunovIIevolution}
\d_\tau Z=\f{1}{10}Z(1-\f{1}{5}\Sigma_+)^{-1}[\f{1}{3}
(5\Sigma_+-1)^2+
(-3N_1^2+\Sigma_+^2-15\Sigma_++4)(1-Q)]
\end{equation}
The restriction of $Z$ to the set $q=\infty$ is non-decreasing along solutions 
as a consequence of (\ref{liapunovIIevolution}). Moreover, it is strictly 
increasing unless
$\Sigma_+=\f{1}{5}$. When $\Sigma_+=\f{1}{5}$ it follows from 
(\ref{bianchiA}) 
that $\dot\Sigma_+
\ne 0$ unless $N_1=\f{2\sqrt2}{5}$. Thus apart from the stationary 
solution at
the point $P_7$ with coordinates $(\infty,\f{2\sqrt2}{5},\f{1}{5})$, 
the function 
$Z$ is strictly increasing along any solution with $q=\infty$. It follows that 
$P_7$ is the $\omega$-limit point of all solutions. The function $Z$ attains 
its minimum precisely on the boundary of the region where the system is 
defined and hence the $\alpha$-limit set of any solution is contained in this 
boundary. The only stationary points on the boundary are $P_5$ and $P_6$.
{}From Lemma 4.2 it follows that both are saddle points of this system. ($P_5$
is degenerate while $P_6$ is non-degenerate.) This suffices to show, using 
Lemma A1 and Lemma A2 of the appendix, that the $\alpha$-limit set consists 
of the entire boundary.

\vskip 10pt\noindent
{\bf Proof of Theorem 4.2} By Lemma 4.1 the $\omega$-limit set of any 
solution consists of points with $q=\infty$. Then Lemma 4.4 shows that either
$P_7$ belongs to the $\omega$-limit set or that the $\omega$-limit set 
consists entirely of points with $q=\infty$ and $N_1=0$ or 
$-\f{1}{2} N_1^2+\f{2}{3}(1-\Sigma_+^2)=0$. A calculation of the 
linearization
of (\ref{bianchiA}) around $P_7$ shows that this point is a hyperbolic sink. 
Hence if
$P_7$ belongs to the $\omega$-limit set this set consists of $P_7$ alone. 
It remains to rule out the other possibility where the solution has an
$\omega$-limit point on the boundary of the intersection of $S_2$ with 
$q=\infty$. In that case Lemma A1 applied to the point $P_6$ and Lemma A2
applied to the point $P_5$ show that the $\omega$-limit set contains the 
whole of this boundary. It will now be shown using (\ref{liapunovIIevolution}) 
that this leads to a contradiction.
There exist $\delta_1>0$ and $M>0$ such that if $|\Sigma_+
-\f{1}{5}|>\delta_1$ and
$q>M$ the right hand side of (\ref{liapunovIIevolution}) is positive. This is 
because the first
term dominates the second. By reducing $\delta_1$ and increasing $M$ if 
necessary it can be ensured that there exist positive constants $\eta_1$,
$\eta_2$ and $\delta_2$ such that $Z^{-1}\d_\tau Z$ can be bounded below by 
$\eta_1$ as long as $|\Sigma_+-\f{1}{5}|>\delta_1$ and $q>M$ and 
$\dot\Sigma_+>\eta_2$ for $|\Sigma_+-\f{1}{5}|<\delta_1$, 
$|N_1-\f{2\sqrt2}{5}|>\delta_2$ and $q>M$. Finally, given $\eta_3>0$ there 
exists 
$\delta_3>0$ so that $\dot\Sigma<\eta_3$ for $|\Sigma|>1-\delta_3$ and $q>M$.
At sufficiently late times the solution lies in the region $q>M$. Moreover,
under the present assumption on the $\omega$-limit set it cannot enter the
neighbourhood of $P_7$ defined by $\delta_1$ and $\delta_2$. Each time it
crosses the strip defined by $|\Sigma_+-\f{1}{5}|\le \delta_1$ at a 
sufficiently
late time it must enter the
region $\Sigma_+>1-\delta_3$ before it can return to the strip. It must spend
a long time in the region $\Sigma_+>1-\delta_3$ (due to the smallness of 
$\eta_3$). This time can be bounded below by $C\eta_3^{-1}$ for a constant 
$C>0$. During that time $\log Z$ must increase by at least $C(\eta_1/\eta_3)$.
On the other hand $\log Z$ can only decrease while it is in 
the strip. It stays there for a time at most $\delta_1/\eta_2$ and can decrease
by at most $C\delta_1/\eta_2$. Thus the net change of $Z$ for each time it 
enters the strip is at least $C(\eta_1/\eta_3-\delta_1/\eta_2)$. If $\eta_3$
is chosen small enough this will be bounded below by a positive quantity.
Since the solution must, under the given assumptions, enter the strip 
infinitely often, this gives a contradiction. The proof of the statement 
about the $\omega$-limit set is now complete.
 
Suppose that the $\alpha$-limit set contains a point with $q=0$ and 
$N_1\ne 0$. Then by Lemma 4.3 it contains $P_2$ and $P_3$ or $P_4$. 
Lemma 4.1 then shows that at least one of $P_1,\ldots,P_6$ is contained in 
the $\alpha$-limit set. If the $\alpha$-limit set contains $P_1$ then either 
the solution lies in the unstable manifold of $P_1$, which gives case (i) of 
the theorem, or the $\alpha$-limit set contains points on that unstable 
manifold other than $P_1$ itself. But since these satisfy neither $N_1=0$ or 
$q=0$ this is a contradiction. If the $\alpha$-limit set contained points 
with $N_1=0$ with $q$ finite and $|\Sigma_+|<1$ it would contain $P_1$, 
leading once again to a contradiction. If it contains $P_4$ it follows from
Lemma A1 and what has just been said that it must contain either $P_3$ or 
$P_6$. It must also contain $P_2$. However, if it contained $P_3$ it would,
by another application of the same lemma contain $P_1$, a contradiction.
On the other hand, if it contains $P_2$ it must contain $P_4$ and $P_5$.
If it contains $P_5$ it must contain $P_6$ and vice versa, by Lemma A1 and
Lemma A2. On the other hand, Lemma A1 shows that if the $\alpha$-limit set
contains $P_5$ or $P_6$ it must contain $P_2$ or $P_4$. It also follows
from these applications of the lemmas of the appendix that the relevant 
connecting orbits are contained in the $\alpha$-limit sets.

\vskip 10pt\noindent
Now a spacetime corresponding to the point $P_7$ will be 
determined (in figure 2, this space-time follows a straight line at
constant $(\Sigma_+,N_1)$ into $P_7$.). {}From 
equation (\ref{meanevolutionsurf}) it follows that 
$\tr k=H_0 e^{-\f{3}{5} \tau}$. Putting this in 
the equation relating $t$ and $\tau$ shows that $\tr k=-\f{5}{3} t^{-1}$. 
Putting this in the third equation of (\ref{dimensionless}) gives 
$b=b_0 t^{2/3}$. Equation (\ref{meancurv})
implies that $a=a_0 t^{1/3}$. Finally, the second equation of  
(\ref{dimensionless}) leads 
to the relation $a_0=\f{2\sqrt2}{5} b_0^2$. Choosing an explicit 
representation
of a Bianchi type II frame leads to the metric:
\begin{equation}\label{explicit}
ds^2=-dt^2+ \f{8}{9} B^2t^{2/3} (dx+zdy)^2+Bt^{4/3}(dy^2+dz^2)
\end{equation}
where $B$ is a constant. This metric is invariant under the homothety 
$t\mapsto At$, $x\mapsto A^{2/3}x$, $y\mapsto A^{1/3}y$, 
$z\mapsto A^{1/3}z$. It follows that $t\d/\d t+\f{2}{3}x\d/\d x
+\f{1}{3}(y\d/\d y
+z\d/\d z)$ 
is a homothetic vector field and that this metric is self-similar. It
satisfies the Einstein equations with an energy-momentum tensor whose only
non-vanishing components are $\rho$ and $T_{11}$. These two are equal and 
are proportional to $t^{-2}$. This can be interpreted as a distributional
solution of the Einstein-Vlasov equations with massless particles where the 
distribution function is of the form $f(p_1,p_2,p_3)=f_1(p_1)\delta(p_2)
\delta(p_3)$. (Note that a distributional $f$ of this kind defines a
dynamical system just as a smooth $f$ does so that the solution can be
represented in figure 2.)
The exact form of the function $f_1$ is unimportant. Only
the integrals $\int f_1(p_1)p_1 dp_1$ and $\int f_1(p_1)p_1^2 dp_1$
influence the energy-momentum tensor. Related to this fact is that the same 
spacetime can be interpreted as a solution of the Einstein equations coupled 
to two streams of null dust moving in opposite senses in the 
$x^1$-direction. This corresponds to choosing 
$f_1=(\delta(p_1)+\delta(-p_1))$ instead of a smooth 
function. The sum of two Dirac measures is necessary to preserve the 
reflection symmetry. This spacetime has previously been considered by
Dunn and Tupper\cite{dunn80} in the context of cosmological models with 
electromagnetic
fields, although it had to be rejected for their purposes since no
consistent electromagnetic field existed.

The monotone function $Z$ which plays a crucial role in the proof of Theorem
4.2 is rather complicated and so is unlikely to be found by trial and error.
We found it by means of a Hamiltonian formulation of the equations for 
$q=\infty$. Once the function was found for $q=\infty$ it was extended so
as to be independent of $q$. In developing the Hamiltonian formulation we
followed the treatment of Uggla in chapter 10 of \cite{wainwright97}. A key 
point is that
the energy density of a distributional solution of the Einstein-Vlasov system
where the pressure is concentrated in one direction can be related in a simple 
way to $q$. The function $Z$ is the Hamiltonian for the (time dependent)
Hamiltonian system. It was also the construction of $Z$ which led us to 
discover the self-similar solution corresponding to the point $P_7$.

The picture obtained in Theorem 4.2 is quite different from that seen in
LRS Bianchi type II solutions with a perfect fluid with linear equation of
state as matter model (see \cite{wainwright97}, chapter 6).
There generic solutions are approximated near the
singularity by a vacuum solution (the type II NUT solution) and there is
no oscillatory behaviour. In the expanding direction the fluid solutions
are also all asymptotic to a self-similar solution (the Collins-Stewart
solution) but this solution has a different ratio of shear to expansion
than the solution corresponding to the point $P_7$. Moreover the pressure
is highly anisotropic in the latter solution.

\section{Kantowski-Sachs and Bianchi type III models}

In this section information will be obtained on Kantowski-Sachs models and 
models of Bianchi type III which is as complete as that obtained on models
of Bianchi type I in section 3.

\vskip 10pt\noindent
{\bf Theorem 5.1} If a smooth non-vacuum reflection symmetric  
Kantowski-Sachs type solution of the Einstein-Vlasov equations for massless 
particles 
is represented as a solution of (\ref{surfacesymm}) with $\epsilon=1$ then for 
$\tau\to -\infty$ either

\noindent
(i) it converges to $P_1$ 
\next
(ii) it converges to the point $P_3$ and it belongs to the unstable manifold 
of $P_3$ or
\next
(iii) it converges to $P_4$
\next
All of these cases occur, and (iii) is the generic case in the sense that it
occurs for an open dense set of initial data.

\noindent
{\bf Proof} The inequality $\d_\tau B\ge\f{1}{4} B$ shows that $B$ decreases
towards the past. It follows that as $\tau$ decreases the solution remains
in a compact set and hence that the solution exists for all sufficiently
negative $\tau$. Using the inequality again shows that $B\to 0$ exponentially
as $\tau\to -\infty$ and the $\alpha$-limit set lies in the set $B=0$. The 
latter can be identified with the Bianchi I system. The $\alpha$-limit set
contains the image of a solution of the Bianchi I system and hence, by 
Theorem 3.1 contains either $P_1$ or some point of the boundary of the 
Bianchi I system. Each of the stationary points $P_1,\dots,P_6$, considered 
as stationary points of (\ref{surfacesymm}), has a linearization which differs 
from its 
linearization within the Bianchi I system by the addition of an extra 
eigenvector with a positive eigenvalue. It can be concluded from this
that $P_4$ is a hyperbolic source. Moreover, by Lemma A1 and Lemma A2, if any 
point of the boundary other than $P_3$ lies in the $\alpha$-limit set, then 
$P_4$ must also lie in the $\alpha$-limit set. Hence in this case the 
$\alpha$-limit set consists of $P_4$ alone. Moreover, if $P_3$ lies in the 
$\alpha$-limit set then the solution must lie on its unstable manifold. The 
only remaining possibility is that the $\alpha$-limit set consists of $P_1$ 
alone, and that the solution lies on the unstable manifold of $P_1$.    

\vskip 10pt\noindent
No statement is made here about the behaviour as $\tau\to\infty$. In fact 
any solution of (\ref{surfacesymm}) with $\epsilon=1$ tends to infinity in 
finite time.
However this is not a problem from the point of view of understanding the 
spacetime. It is known that the Kantowski-Sachs models recollapse 
\cite{burnett91}.
Thus there is no infinitely expanding phase and a final singularity 
looks like the time reverse of an initial singularity. One interesting 
question which we do not attempt to tackle here is whether there is an 
interesting correlation between the behaviour near the initial and final 
singularities. For each individual singularity the picture is essentially 
identical to that seen in the singularity of Bianchi I models. The system 
for a radiation fluid can be analysed in the same way, reducing the dynamics 
near the singularity to that of the corresponding Bianchi I system. The 
differences between radiation fluid and kinetic models are similar in both 
cases.

\vskip 10pt\noindent
{\bf Theorem 5.2} If a smooth non-vacuum reflection symmetric LRS solution of 
Bianchi type III of the Einstein-Vlasov equations for massless particles is 
represented as a solution of (\ref{surfacesymm}) with $\epsilon=-1$ then for 
$\tau\to \infty$ it converges to the the point $P_9$ with coordinates
$(\infty,\f{1}{2},\f{1}{2})$ and for $\tau\to -\infty$ either

\noindent
(i) it converges to $P_1$ or
\next
(ii) it converges to the point $P_3$ and it belongs to the unstable manifold 
of $P_3$ or
\next
(iii) it converges to $P_4$
\next
All of these cases occur, and (iii) is the generic case in the sense that it
occurs for an open dense set of initial data.

\noindent
{\bf Proof} The inequality (\ref{physicalsurf}) with $\epsilon=-1$ implies 
that a solution of (\ref{surfacesymm}) of Bianchi type III remains in a 
compact set 
and hence exists 
globally in $\tau$. The quantity $\dot B$ is positive in the region
where $B^2<\f{1}{4}+\f{1}{12}(1-2\Sigma_+)^2$. Call this region $G$. 
In the 
complement of the closure of $G$ the inequality $\dot B<0$ holds. Thus any
stationary point with $B>0$ must lie on the boundary of $G$. A stationary 
point with a finite non-zero value of $q$ must satisfy $\Sigma_+=0$ and this
implies that $\dot\Sigma=\f{1}{3}$, a contradiction. Thus the only 
stationary
points occur for $B=0$ (these are the well-known Bianchi type I stationary
points), $q=0$ or $q=\infty$. In fact the only stationary points which are 
not of type I are those with coordinates $(0,\f{1}{2},\f{1}{2})$ and 
$(\infty,\f{1}{2},
\f{1}{2})$. Call these $P_8$ and $P_9$ respectively (see figure
3). \\

\noindent The boundary of $G$ is connected and so 
$\dot\Sigma$ has a constant sign there. Checking at one point shows that 
this sign is positive. As a consequence, a solution can never leave $G$
as $\tau$ increases or enter $G$ as $\tau$ decreases. A solution which lies
on the boundary of $G$ at some time (with $q$ non-zero and finite) must
immediately enter $G$ to the future and enter the interior of its complement
to the past. Consider now the behaviour of a given solution as $\tau$
decreases. If it stayed in $G$ for ever then $B$ would have to increase
as $\tau$ decreases. On the other hand, any $\alpha$-limit point would have 
to be in the boundary of $G$ due to the monotonicity properties of $B$. This
is not consistent. Thus as $\tau$ decreases the solution must reach the 
boundary of $G$ and, as a consequence the interior of the complement of $G$.
In the latter region $B$ is strictly monotone and so the $\alpha$-limit set 
must be contained in $B=0$. Then the same analysis as in the proof of Theorem 
5.1 shows that the solution belongs to one of the cases (i)-(iii) of the 
theorem. Next consider the behaviour as $\tau$ increases. As $\tau$ tends to
infinity the solution must tend to the boundary of $G$. If it stays in the 
interior of the complement of $G$ then it must tend to the boundary of $G$
as $\tau$ tends to infinity and, more precisely, to one of the points $P_8$ 
or $P_9$. Since $\Sigma_+$ is positive at these points, $q\to\infty$ and so
only $P_9$ is possible. Now suppose that the solution does meet the boundary
of $G$ and hence enters $G$ itself. Then it remains in $G$ and $B$ is once
again strictly monotone. As before, it can be concluded that the solution 
converges to $P_9$ as $\tau\to\infty$.
\begin{figure}
\begin{center}
\includegraphics[width=10cm,height=8cm,angle=270]{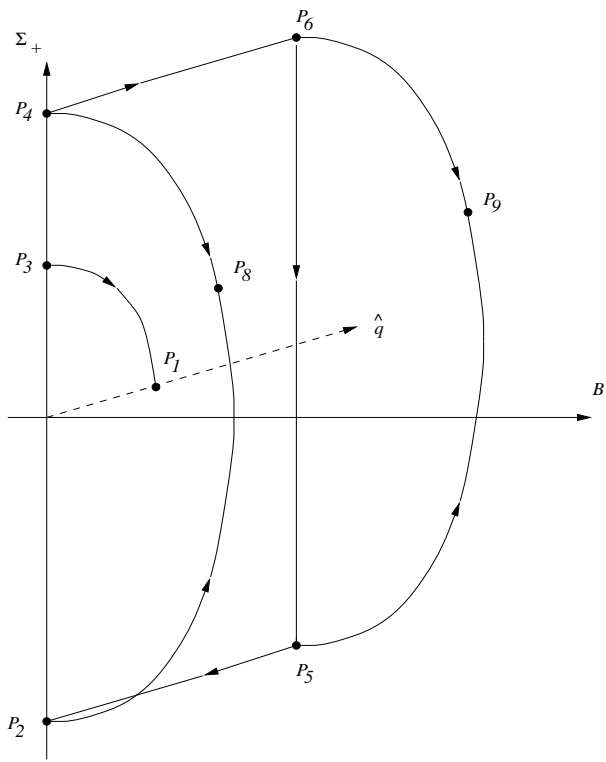}
\end{center}
\label{Figure 3}
\caption{The $(\hat{q},\Sigma_+,B)$ space and the fixed points for Bianchi
type III.}
\end{figure}
\vskip 10pt\noindent
The point $P_9$ corresponds to a self-similar solution of the vacuum Einstein
equations much as does $P_7$ (this time the trajectory is the
horizontal straight line in figure 3 from $P_8$ to $P_9$). This is the Bianchi III form of flat space (see p. 193 of 
\cite{wainwright97}).
Once again the nature of the initial singularity is similar to that in 
solutions of type I. On the other hand the final singularity is 
qualitatively different from any we have seen so far. In this case the
solution is approximated at large times by a vacuum solution in the sense
that the dimensionless quantity $\rho/(\tr k)^2$ tends to zero as $t\to\infty$.
A very similar analysis applies to the system for a radiation fluid.
LRS Bianchi type III fluid solutions with equation of state $p=\f{1}{3}
\rho$ 
behave like solutions of Bianchi type I near the initial singularity. They 
are approximated at large times by the same vacuum solution as in the case of 
kinetic theory. The approach of \cite{hewitt93} should allow similar 
statements to be 
proved for other fluids with a linear equation of state, but this does not 
seem to have been worked out explicitly in the literature.

\section{Conclusions}

The above theorems show that solutions of the Einstein-Vlasov equations
with high symmetry exhibit a wide variety of asymptotic behaviour near a
singularity and in a phase of unlimited expansion. They can have a point 
singularity, barrel singularity or cigar singularity or they can show 
oscillatory behaviour near a singularity. In an expanding phase they can
resemble a fluid solution (Bianchi type I and VII${}_0$), a vacuum solution 
(Bianchi type III) or a solution of the Einstein equations with null dust
(Bianchi type II).
There are notable differences in comparison with a fluid model, and this
includes the radiation fluid, which is often used as an effective model of
massless particles in cosmology. The most striking qualitative difference
is the appearance of oscillatory behaviour in type II solutions. It is 
interesting to compare this with the analysis of spacetime singularities
by Belinskii, Khalatnikov and Lifschitz \cite{belinskii82}. They do not say 
precisely what 
they assume about matter but it seems that they do assume, at least 
implicitly, that pressures cannot approach the energy density. This
assumption is not necessarily satisfied in a kinetic description. The mean
pressure cannot exceed one third of the energy density but if it all 
concentrates in one direction the pressure in that direction can approach
the energy density. This leads to a source of oscillations beyond those
taken into account in \cite{belinskii82}.

While the oscillatory behaviour of cosmological models near a singularity
has often been observed numerically and explained heuristically, it has
rarely been captured in rigorous theorems. To our knowledge the only
example where this had been done previous to Theorem 4.2 of this paper
is in a class of solutions of the Einstein-Maxwell equations of Bianchi type 
VI${}_0$ analysed in \cite{leblanc95}.

The results of this paper concern only massless particles. One may ask
what would change in the results if the case of massive particles is
considered. In one case the answer is known, namely in Bianchi type I.
There the solution approaches a dust solution in the expanding phase.
It is reasonable to expect that this happens more generally. As the model
expands in all directions the pressures should become negligible with 
respect to the energy density, leading to a dust-like situation.
However, the techniques necessary to prove this are not yet known.
Near the initial singularity, the equations for massive particles look
like those for massless particles and it may be conjectured that the 
behaviour near the singularity is similar in both cases. Unfortunately
that has also not yet been proved.

It is interesting to note that matter seems to have the effect of making
the evolution of the geometry under the Einstein equations less extreme
in a phase of unlimited expansion.
In Bianchi type I the vacuum solutions (Kasner solutions) are such that 
some spatial direction is contracting or unchanging in the expanding 
time direction (the time direction in which the volume is increasing). This 
is no longer the case when perfect fluid or kinetic matter is added, since 
then the model isotropizes. In type II there is no complete isotropization 
but it is still the case that with fluid or kinetic matter all directions are 
eventually expanding, in contrast to the vacuum case. The type III case is 
borderline, since there solutions with collisionless matter are asymptotic, 
in the sense of the variables used in this paper, to a vacuum solution in the 
expanding time direction. The vacuum solution is such that the scale factor
$b$ is time independent. In the other LRS Bianchi type III vacuum spacetimes
this scale factor is asymptotically constant as $t\to\infty$ and for a 
radiation fluid this is also the case (cf. \cite{wainwright97} p. 203). On 
the other hand 
for dust models which also converge to the same vacuum model in terms of the
Wainwright-Hsu variables, this scale factor grows without bound, although
much more slowly than the other scale factors (\cite{wainwright97}, p. 202). 
It is difficult
to decide what happens in the case of collisionless matter with massless
particles, since the point $P_9$ is a degenerate stationary point of the 
system (\ref{surfacesymm}). In the corresponding system for a radiation fluid 
the point 
with these coordinates is also a stationary point but is non-degenerate.

For the Einstein-Vlasov equations with 
massless particles the LRS reflection symmetric solutions of Bianchi
types I, II, III, VII${}_0$ and Kantowski-Sachs type have now been analysed
as far as to give a full description of their general behaviour near the
singularity and in a phase of unlimited expansion. 
There are still plenty of open questions related to this. What
happens with LRS solutions of types VIII and IX? What happens if reflection 
symmetry is dropped? Does this lead to a new kind of oscillatory behaviour?
What happens if the LRS condition is dropped? (This is still open even in
the Bianchi I case.) Can the Hamiltonian formulation of the equations, which
played an important role at one point in our arguments, usefully be applied
in some of these more general cases? Answers to these questions could help 
to deepen our understanding of the dynamics of solutions of the Einstein
equations with matter in general.

\vskip 10pt\noindent
{\it Acknowledgements} We wish to thank Malcolm MacCallum for his comments
on the exact solution in section 4. Paul Tod gratefully acknowledges the 
hospitality and financial support of the Max-Planck-Institut f\"ur 
Gravitationsphysik while this work was being done.

\section*{Appendix: Some background on dynamical systems}

First some terminology will be introduced. We use the phrase \lq dynamical 
system\rq\ as a synonym for \lq system of ordinary differential equations\rq. 
The difference between the two is then only one of point of view. A {\it 
stationary point} of a dynamical system is a time-independent solution. An 
{\it orbit} of a dynamical system is the image of a solution. A point $x_*$
is an $\alpha$-limit point of a solution $x(t)$ if there is a sequence of
times $t_n$ with $t_n\to -\infty$ such that $x(t_n)\to x_*$. The set of all
$\alpha$-limit points of a solution is called its $\alpha$-limit set. The
analogous notions of $\omega$-limit point and $\omega$-limit set are obtained
by replacing $t$ by $-t$ in these definitions. Basic properties are that the
$\alpha$-limit set is closed and that, if the solution remains in a compact
set as $t\to -\infty$, it is connected. If $x_*$ is a point of the 
$\alpha$-limit set of an orbit then the orbit through $x_*$ lies in the 
$\alpha$-limit set of the original orbit. Analogous statements hold for the
$\omega$-limit set. For details and proofs see e.g. \cite{hartman82}, 
chapter 7.

If $x_0$ is a stationary point of a dynamical system we can linearize the 
system about $x_0$. The linearized system is of the form 
$d\tilde x/dt=A\tilde x$ for a matrix $A$. Associated to $A$ is a direct sum
decomposition $E_1\oplus E_2\oplus E_3$ where the vector spaces $E_1$, $E_2$
and $E_3$ are spanned by generalized eigenvectors of $A$ corresponding to
eigenvalues with positive, zero and negative real parts, respectively. These
spaces are called the unstable, centre and stable subspaces. For each of 
these three subspaces there is a manifold which is tangent to the 
corresponding subspace at $x_0$ and is left invariant by the
dynamical system. These manifolds are called the unstable, centre, and 
stable manifolds respectively. The unstable and stable manifolds are 
unique while the centre manifold need not be. For details see the appendix 
of \cite{abraham67}.

The behaviour of solutions of a dynamical system near a stationary point
is described by the reduction theorem.

\noindent
{\bf Theorem A1} (Reduction theorem) Let $x_0$ be a stationary point of 
a $C^1$ dynamical system. Then the system is topologically equivalent near 
$x_0$ to the Cartesian product of a standard saddle with the restriction of 
the flow to any centre manifold.

\vskip 10pt\noindent
This theorem is proved in \cite{kirchgraber90}. Topological equivalence means 
that there is a
homeomorphism which takes one system to the other. A standard saddle is the
dynamical system on $\R^{n_1+n_2}$ given by $dy/dt=y$, $dz/dt=-z$, where 
$y\in\R^{n_1}$ and $z\in\R^{n_2}$. The special case (hyperbolic case) where 
the centre manifold is trivial is the Hartman-Grobman theorem 
\cite{hartman82}.

The next result is intuitively rather obvious, but since we do not
know a published proof we will provide one here.

\noindent
{\bf Lemma  A1} Let $p$ be a hyperbolic stationary point of a dynamical 
system which belongs to the $\alpha$-limit set of a given orbit. Then either 
each neighbourhood of $p$ contains a segment of the orbit which is contained 
in the unstable manifold of $p$, or the $\alpha$-limit set contains a 
point of the stable manifold of $p$ other than $p$ itself. The analogous
statement with the roles of the stable and unstable manifolds interchanged
also holds. 

\noindent
{\bf Proof} By the reduction theorem we can assume that in a neighbourhood 
of $p$ the system takes the form:
\begin{equation}\label{alinearized}
dx/dt=x,\ \ \ dy/dt=-y
\end{equation}
with solution
\begin{equation}\label{alinearizedsol}
x=Ae^t,\ \ \ y=Be^{-t}
\end{equation}
The unstable and stable manifolds are given by $y=0$ and $x=0$ respectively.
Suppose that there is a neighbourhood of $p$ where there is no 
segment of the orbit contained in the unstable manifold. Then there exists 
a sequence of points $p_n$ on the orbit with non-vanishing $y$ coordinate 
which converges to $p$. If we denote the coordinates of corresponding segments
of the solution by $(x_n,y_n)$, then $y_n=B_ne^{-t}$ for some $B_n\ne 0$.
Consider now a coordinate closed ball contained in a neighbourhood of $p$
where the reduction can be carried out. As $t$ decreases each of the solutions
$(x_n,y_n)$ must leave this ball and so must, in particular contain a
point of the boundary sphere. Call the resulting sequence of points of the 
sphere $q_n$. By compactness $q_n$ has a subsequence converging to a point 
$q$. The point $q$ belongs to the $\alpha$-limit set. Now $x_n=A_ne^t$ for
a sequence with $A_n\to 0$. Hence the $x$ coordinate of $q$ is zero and $q$
belongs to the stable manifold of $p$. The proof in the case that the roles
of the stable and unstable manifolds are interchanged is very similar, using
the points where the solution exits the ball in the positive time direction.

\vskip 10pt\noindent
The following variant of Lemma A1 allows a centre manifold of a certain
type.

\noindent
{\bf Lemma A2} Let $p$ be a stationary point of a dynamical system 
which belongs to the $\alpha$-limit set of a given orbit. Suppose that the
centre manifold is one-dimensional and that there is a punctured
neighbourhood of $p$ in the centre manifold which contains no stationary
points and such that the solutions on the centre manifold approach $p$
as $t\to\infty$ on one side of $p$ and as $t\to -\infty$ on the other
side. Suppose further that the stable manifold is trivial. The boundary 
between points on orbits which converge to $p$ while staying in a small
neighbourhood of $p$ as $t\to -\infty$ and points on orbits which do not is 
the unstable manifold. The analogue of Lemma A1 holds, where the stable 
manifold is replaced by the half of the centre manifold on one side of the 
unstable manifold. This half of the centre manifold is unique.

\noindent
{\bf Proof} By the reduction theorem we can assume that in a neighbourhood 
of $p$ the system takes the form:
\begin{equation}\label{reduction}
dx/dt=F(x),\ \ \ dy/dt=y
\end{equation}
for some function $F$ which vanishes together with its derivative at the 
origin, and is positive otherwise. The boundary hypersurface is given by
$x=0$. The half of the centre manifold referred to in the statement of the
theorem corresponds to $x<0$ and $y=0$. In the half-plane $x<0$ the system
is topologically equivalent to a hyperbolic saddle and so it is possible 
to obtain the conclusion as in the proof of Lemma A1. 

Next we state the Poincar\'e-Bendixson theorem. The form of this theorem 
which we will use is the following (cf. \cite{hartman82}, p. 151):

\noindent
{\bf Theorem A2} (Poincar\'e-Bendixson) Let $U$ be an open subset of 
$\R^2$ and consider a dynamical system on $U$ without stationary points.
Let $x(t)$ be a solution which exists globally and remains in a compact 
subset of $U$ as $t\to -\infty$. Then the $\alpha$-limit set of the given 
solution is a periodic orbit. 

\vskip 10pt\noindent
The analogous statement holds for the the $\omega$-limit set. A periodic 
orbit is, of course, just the image of a periodic solution.

\end{document}